\documentclass[dvips]{acta}
\usepackage{supertabular,lscape,epsfig}
\usepackage{amssymb}
\usepackage{amsmath}
\DeclareSymbolFont{ppa}{OT1}{ppl}{m}{it}
\DeclareMathSymbol{\vv}{\mathalpha}{ppa}{'166}

\newfont{\hb}{rphvb at 10pt}
\newfont{\hbo}{rphvbo at 10pt}
\newfont{\bitt}{rptmbi at 12pt}
\newfont{\bits}{rptmbi at 11pt}

\SetPages{1}{23}

\SetVol{61}{2011}

\begin{document}

\newcommand{\TabApp}[2]{\begin{center}\parbox[t]{#1}{\centerline{
  {\bf Appendix}}
  \vskip2mm
  \centerline{\small {\spaceskip 2pt plus 1pt minus 1pt T a b l e}
  \refstepcounter{table}\thetable}
  \vskip2mm
  \centerline{\footnotesize #2}}
  \vskip3mm
\end{center}}

\newcommand{\TabCapp}[2]{\begin{center}\parbox[t]{#1}{\centerline{
  \small {\spaceskip 2pt plus 1pt minus 1pt T a b l e}
  \refstepcounter{table}\thetable}
  \vskip2mm
  \centerline{\footnotesize #2}}
  \vskip3mm
\end{center}}

\newcommand{\TTabCap}[3]{\begin{center}\parbox[t]{#1}{\centerline{
  \small {\spaceskip 2pt plus 1pt minus 1pt T a b l e}
  \refstepcounter{table}\thetable}
  \vskip2mm
  \centerline{\footnotesize #2}
  \centerline{\footnotesize #3}}
  \vskip1mm
\end{center}}

\newcommand{\MakeTableApp}[4]{\begin{table}[p]\TabApp{#2}{#3}
  \begin{center} \TableFont \begin{tabular}{#1} #4 
  \end{tabular}\end{center}\end{table}}

\newcommand{\MakeTableSepp}[4]{\begin{table}[p]\TabCapp{#2}{#3}
  \begin{center} \TableFont \begin{tabular}{#1} #4 
  \end{tabular}\end{center}\end{table}}

\newcommand{\MakeTableee}[4]{\begin{table}[htb]\TabCapp{#2}{#3}
  \begin{center} \TableFont \begin{tabular}{#1} #4
  \end{tabular}\end{center}\end{table}}

\newcommand{\MakeTablee}[5]{\begin{table}[htb]\TTabCap{#2}{#3}{#4}
  \begin{center} \TableFont \begin{tabular}{#1} #5 
  \end{tabular}\end{center}\end{table}}

\newfont{\bb}{ptmbi8t at 12pt}
\newfont{\bbb}{cmbxti10}
\newfont{\bbbb}{cmbxti10 at 9pt}
\newcommand{\uprule}{\rule{0pt}{2.5ex}}
\newcommand{\douprule}{\rule[-2ex]{0pt}{4.5ex}}
\newcommand{\dorule}{\rule[-2ex]{0pt}{2ex}}
\def\thefootnote{\fnsymbol{footnote}}
\begin{Titlepage}

\Title{The Optical Gravitational Lensing Experiment.\\
The OGLE-III Catalog of Variable Stars.\\
XI.~RR~Lyrae Stars in the Galactic Bulge\footnote{Based on
observations obtained with the 1.3-m Warsaw telescope at the Las Campanas
Observatory of the Carnegie Institution of Washington.}}
\Author{I.~~S~o~s~z~y~ñ~s~k~i$^1$,~~
W.\,A.~~D~z~i~e~m~b~o~w~s~k~i$^1$,~~
A.~~U~d~a~l~s~k~i$^1$,~~
R.~~P~o~l~e~s~k~i$^1$,\\
M.\,K.~~S~z~y~m~a~ñ~s~k~i$^1$,~~
M.~~K~u~b~i~a~k$^1$,~~
G.~~P~i~e~t~r~z~y~ñ~s~k~i$^{1,2}$,\\
£.~~W~y~r~z~y~k~o~w~s~k~i$^{1,3}$,~~
K.~~U~l~a~c~z~y~k$^1$,~~
S.~~K~o~z~³~o~w~s~k~i$^1$
~~and~~ P.~~P~i~e~t~r~u~k~o~w~i~c~z$^1$}
{$^1$Warsaw University Observatory, Al.~Ujazdowskie~4, 00-478~Warszawa, Poland\\
e-mail:
(soszynsk,wd,udalski,rpoleski,msz,mk,pietrzyn,kulaczyk,simkoz,pietruk)@astrouw.edu.pl\\
$^2$ Universidad de Concepción, Departamento de Astronomia, Casilla 160--C, Concepción, Chile\\
$^3$ Institute of Astronomy, University of Cambridge, Madingley Road, Cambridge CB3 0HA, UK\\
e-mail: wyrzykow@ast.cam.ac.uk}
\Received{March 18, 2011}
\vspace*{-6pt}
\end{Titlepage}
\vspace*{-9pt}
\Abstract{The eleventh part of the OGLE-III Catalog of Variable Stars
(OIII-CVS) contains 16\,836 RR~Lyr stars detected in the OGLE fields toward
the Galactic bulge. The total sample is composed of 11\,756 RR~Lyr stars
pulsating in the fundamental mode (RRab), 4989 overtone pulsators (RRc),
and 91 double-mode (RRd) stars. About 400 RR~Lyr stars are members of the
Sagittarius Dwarf Spheroidal Galaxy. The catalog includes the time-series
photometry collected in the course of the OGLE survey, basic parameters of
the stars, finding charts, and cross-identifications with other catalogs of
RR~Lyr stars toward the Milky Way center.

We notice that some RRd stars in the Galactic bulge show unusually short
periods and small ratio of periods, down to $P_{\rm F}\approx0.35$~days and
$P_{\rm 1O}/P_{\rm F}\approx0.726$. In the Petersen diagram double-mode RR~Lyr
stars form a parabola-like structure, which connects shorter- and
longer-period RRd stars. We show that the unique properties of the bulge
RRd stars may be explained by allowing for the wide range of the metal
abundance extending up to [Fe/H]$\approx-0.36$.

We report the discovery of an RR~Lyr star with additional eclipsing
variability with the orbital period of 15.2447~days. Some statistical
features of the RR~Lyr sample are presented. We discuss potential
applications of our catalog in studying the structure and history of the
central region of the Galaxy, mapping the interstellar extinction toward
the bulge, studying globular clusters and the Sagittarius Dwarf Galaxy.}
{Stars: variables: RR~Lyrae -- Stars: oscillations (including pulsations)
-- Stars: Population II -- Galaxy: center -- Galaxies: individual:
Sagittarius Dwarf Spheroidal Galaxy}

\Section{Introduction}

RR~Lyrae stars are of particular interest to astronomers for several
reasons. First, they are useful indicators of old, metal-poor population of
stars. Second, they are standard candles, enabling an estimate of their
distances to be made. Third, they are very numerous and bright enough that
they can be easily observed in our and nearby galaxies. Thus, RR~Lyr stars
play an essential role in our understanding of the formation and evolution
of the Galaxy, as well as the internal constitution and evolution of
stars. RR~Lyr stars in the Galactic bulge are an important source of
information on the distance to the center of the Milky Way, the geometry of
the bar and the bulge and the interstellar extinction in these regions. The
properties of these stars give us an insight into the earliest history of
our Galaxy.

First significant sample of RR~Lyr stars close to the central regions of
the Milky Way was discovered by van Gent (1932, 1933). He noticed that
cluster type variables (the historical name of RR~Lyr stars) strongly
concentrate toward the Galaxy center. The fields located closer to the
Galactic center were observed under a survey conducted by the Harvard
Observatory (\eg Swope 1936, 1938). Baade (1946, 1951) observed the
relatively unobscured area today called Baade's Window\footnote{It is
interesting to note that Baade called this region ``van Tulder's pole''.}
which yielded over 100 newly identified RR~Lyr stars (Gaposchkin 1956).

Each of the many other efforts to detect variable stars in the Galactic
bulge (\eg Plaut 1948, 1973, Fokker 1951, Oosterhoff and Horikx 1952,
Oosterhoff \etal 1954, 1967, Ponsen 1955, Oosterhoff and Ponsen 1968,
Hartwick \etal 1981, Blanco 1984) gave additionally from a few to a few
dozen new RR~Lyr stars. As a result, in the early nineties of the twentieth
century about one thousand RR~Lyr variables inhomogeneously distributed
over the Galactic bulge were known.

In the nineties, large and homogeneous samples of variable stars were
published as by-products of microlensing sky surveys. Udalski \etal (1994,
1995ab, 1996, 1997) published a catalog of over 3000 periodic variable
stars in the Galactic bulge detected in the fields covered by the first
phase of the Optical Gravitational Lensing Experiment (OGLE-I). In total,
215 of these stars were classified as RR~Lyr variables. The next stage of
the OGLE project (OGLE-II) resulted in much larger samples of RR~Lyr stars
in the central regions of the Milky Way. Mizerski (2003) detected and
analyzed over 2700 RR~Lyr stars in the bulge. He noticed very high incident
rate of Blazhko stars, and very low percentage of RRd stars. The OGLE-II
data were also used by Collinge \etal (2006) to prepare a catalog of 1888
fundamental-mode RR~Lyr stars (RRab).

Also the MACHO microlensing project observed a numerous sample of RR~Lyr
stars toward the Galactic center (Alcock \etal 1997, 1998). The largest to
date catalog of these variables in the bulge was constructed by Kunder
\etal (2008) on the basis of the MACHO database. Their sample contains 3525
RR~Lyr stars of ab type.

This paper presents a catalog of 16\,836 RR~Lyr stars detected in the
fields toward the Galactic bulge monitored during the third phase of the
OGLE project (OGLE-III). This is the eleventh part of the OGLE-III Catalog
of Variable Stars, and the first part of the Catalog containing variable
stars detected outside the Magellanic Clouds. So far we published, among
others, a huge catalog of almost 25\,000 RR~Lyr stars in the Large
Magellanic Cloud (LMC, Soszyñski \etal 2009, hereafter Paper~I) and a ten
times smaller catalog of this type variables in the Small Magellanic Cloud
(SMC, Soszyñski \etal 2010, hereafter Paper~II).

This paper is structured as follows. Section~2 presents the data and their
reduction. In Section~3, we describe the selection and classification
processes. In Section~4, we describe the catalog itself, and Section~5 is
devoted to the comparison of our sample with other catalogs of RR~Lyr stars
in the Galactic bulge. In Section~6 we discuss possible applications of our
catalog in the studies of the central parts of the Milky Way and
Sagittarius Dwarf Galaxy (hereafter Sgr~dSph). Finally, we summarize our
results and in Section~7.

\Section{Observational Data}
Our observations of the Galactic bulge were obtained at Las Campanas
Observatory with the 1.3-m Warsaw telescope. The observatory is operated by
the Carnegie Institution of Washington. During the OGLE-III project
(2001--2009), the Warsaw telescope was equipped with an eight-chip mosaic
camera covering approximately $35\times35$~arcmin in the sky with the scale
of 0.26~arcsec/pixel. Details of the instrumentation setup can be found in
the paper by Udalski (2003).

The time coverage, as well as a number of points obtained by the OGLE
project in the bulge varies considerably from field to field. Some fields
have been monitored since 1992 and for these fields up to several thousand
points per star have been collected until now. Other fields were observed
for only one or two seasons and only several dozen observations were
collected for them. In this study we used only those OGLE-II and OGLE-III
fields for which at least 30 epochs were gathered. These fields range in
the Galactic coordinates within approximately $|l|<11\arcd$ and
$|b|<7\arcd$ and cover an area of 68.7~square degrees.

Observations were obtained through the {\it I} and {\it V} filters closely
resembling the Johnson-Cousins system. The accuracy of the transformations
from the instrumental to the standard magnitudes is better than 0.02~mag
(Udalski \etal 2008). The vast majority of the observations (from 30 to over
5000 points per star, median: 623) were made with the {\it I}-band filter,
while in the {\it V}-band we obtained from a few to several dozen points.

The time-series photometry attached to this catalog was compiled from the
OGLE-II and OGLE-III observations, so it covers up to 13 years (from March
1997 to May 2009). For individual stars both datasets were tied by shifting
the OGLE-II photometry to agree with the OGLE-III light curves. For 279
RR~Lyr stars exclusively the OGLE-II photometry is available. We also
combined the photometry of stars observed in the overlapping regions of two
or more adjacent fields.

The photometry was obtained with the standard OGLE data reduction pipeline
(Udalski \etal 2008) based on the Difference Image Analysis (DIA, Alard and
Lupton 1998, Wo¼niak 2000). For 40 objects in our catalog there is no {\it
I}-band DIA photometry in the OGLE database, due to saturation or location
close to other bright stars. For these objects we provide the photometry
measured with the {\sc DoPhot} package (Schechter \etal 1993). We flag
these stars in the remarks of the catalog.

\Section{Selection and Classification of RR~Lyr Stars}
\Subsection{Single-Period Variables}
A massive periodicity search was performed for all $3\times10^8$ stars
monitored by the OGLE-III survey in the Galactic bulge. To perform this
time-consuming task, we used supercomputers assembled at the
Interdisciplinary Centre for Mathematical and Computational Modelling (ICM)
of the University of Warsaw. The period-search code {\sc Fnpeaks} (by
Z.~Ko³aczkowski -- private communication) was run on each {\it I}-band
light curve with more than 30 points. Ten the highest peaks in the
periodogram were selected and archived with the corresponding amplitudes
and signal-to-noise ratios. Then, each light curve was prewhitened with the
primary period and the procedure of the period search was repeated on the
residual data.

Before we began selection and classification of variable stars, all light
curves were fitted with a series of Fourier cosine functions, and the
Fourier coefficients $R_{21}$, $\phi_{21}$, $R_{31}$, $\phi_{31}$ (Simon
and Lee 1981) were calculated. We used the positions of stars in the
period--Fourier coefficient planes (Fig.~1) to provisionally divide the
sample into pulsating and other stars. However, the main selection
procedure was based on the visual inspection of the light curves. We
inspected all stars with periods between 0.2 and 1.0~day and amplitudes
larger than a limit that depended on the average brightness of the
star. For the brightest objects the amplitude limit reached 0.01~mag, which
allowed us to select a number of RR~Lyr variables blended by other stars.
\begin{figure}[htb]
\vglue3mm
\hglue-9mm{\includegraphics[width=14cm, bb=10 270 585 745]{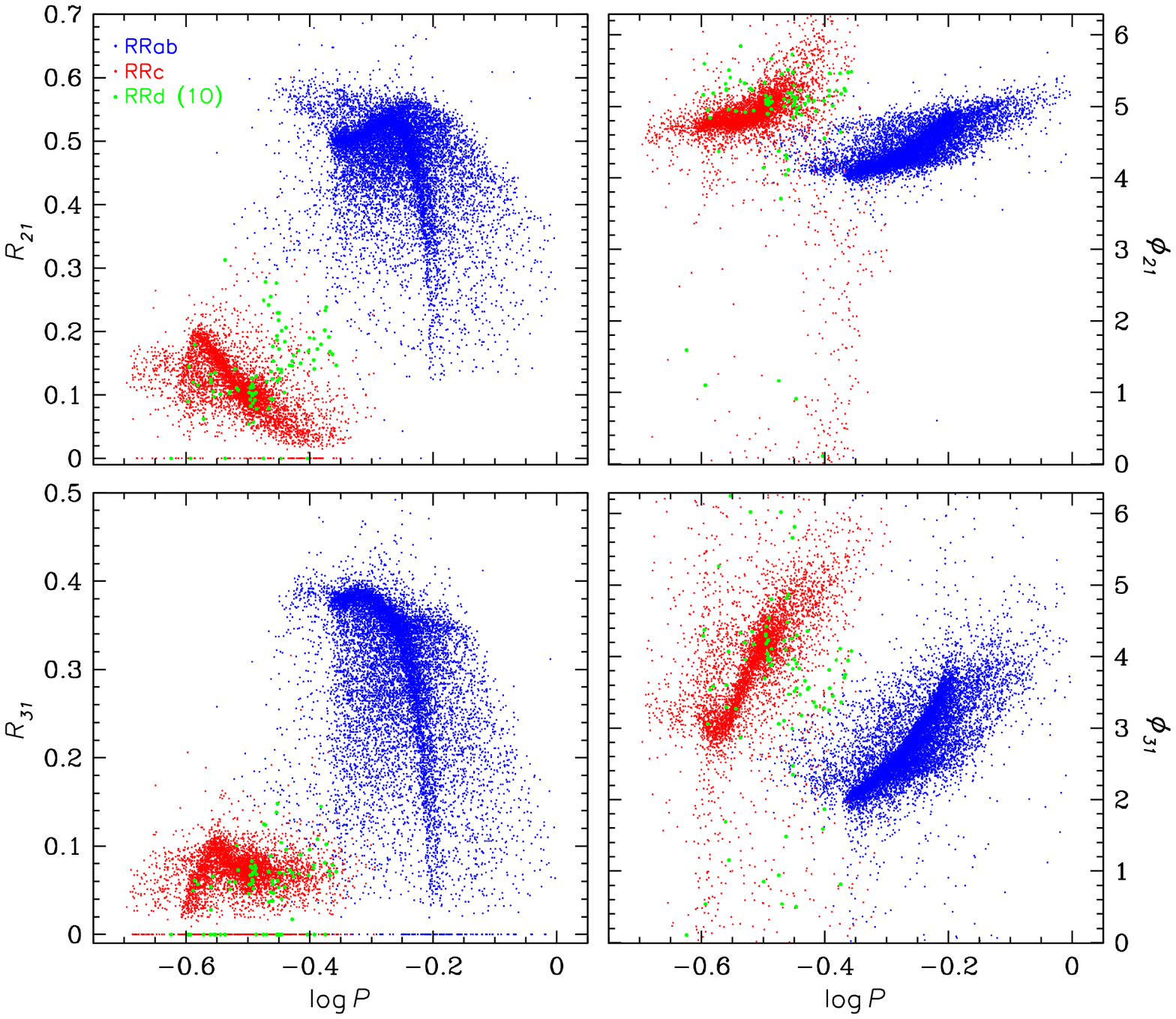}}
\vskip6pt
\FigCap{Parameters $R_{21}$, $\phi_{21}$, $R_{31}$ and $\phi_{31}$ of the
Fourier light curve decomposition (Simon and Lee 1981) plotted against the
logarithm of periods for RR~Lyr stars from our catalog. Blue dots represent
RRab variables, red are RRc stars while green dots show the first overtone
mode of RRd stars.}
\end{figure}
\begin{figure}[p]
\hglue-7mm{\includegraphics[width=15cm, bb=10 410 585 755]{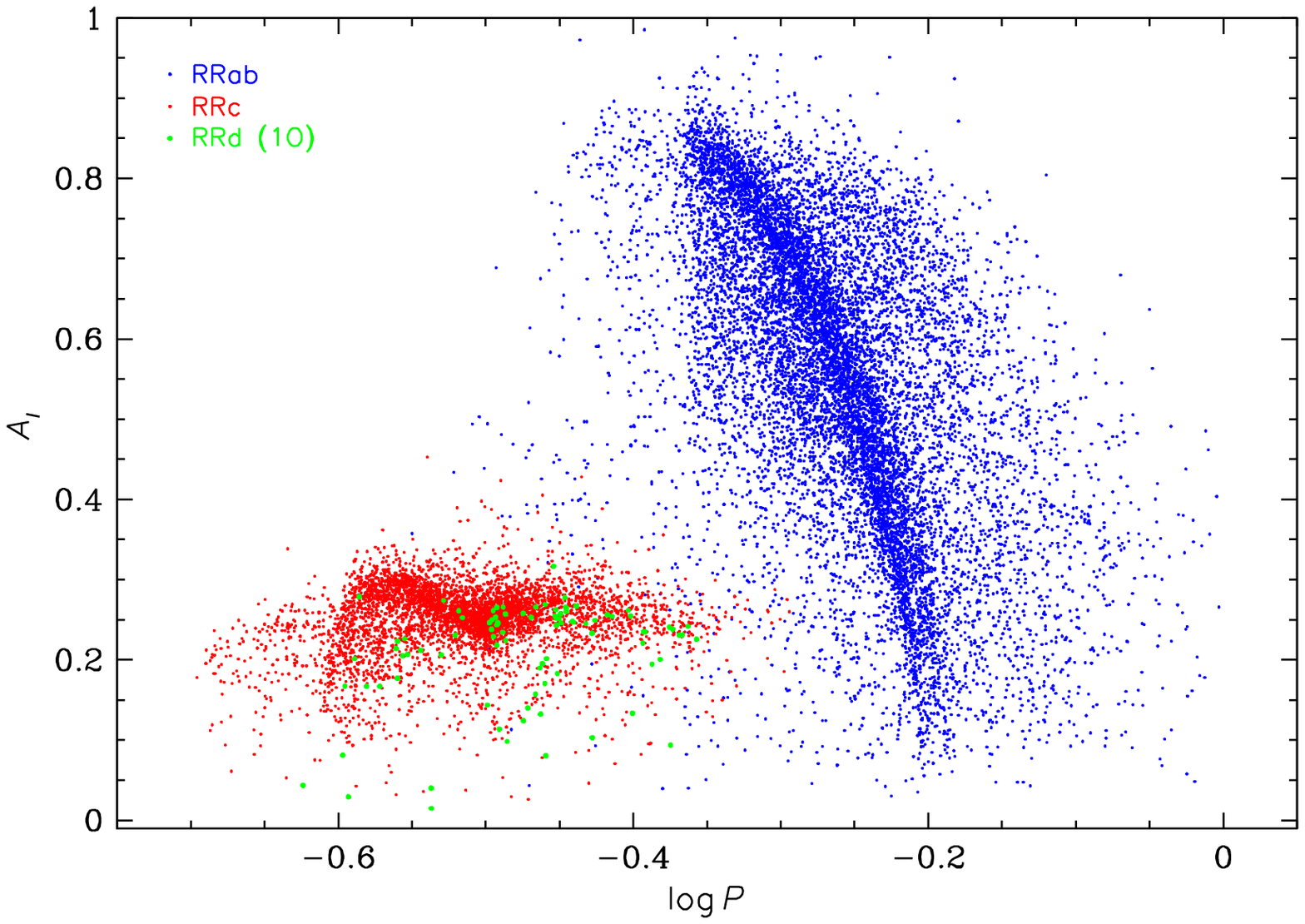}}
\FigCap{Period--amplitude diagram for RR~Lyr stars toward the Galactic
bulge. Different colors represent the same type of stars as in Fig.~1.}
\vskip3mm
\hglue-7mm{\includegraphics[width=15cm, bb=10 410 585 755]{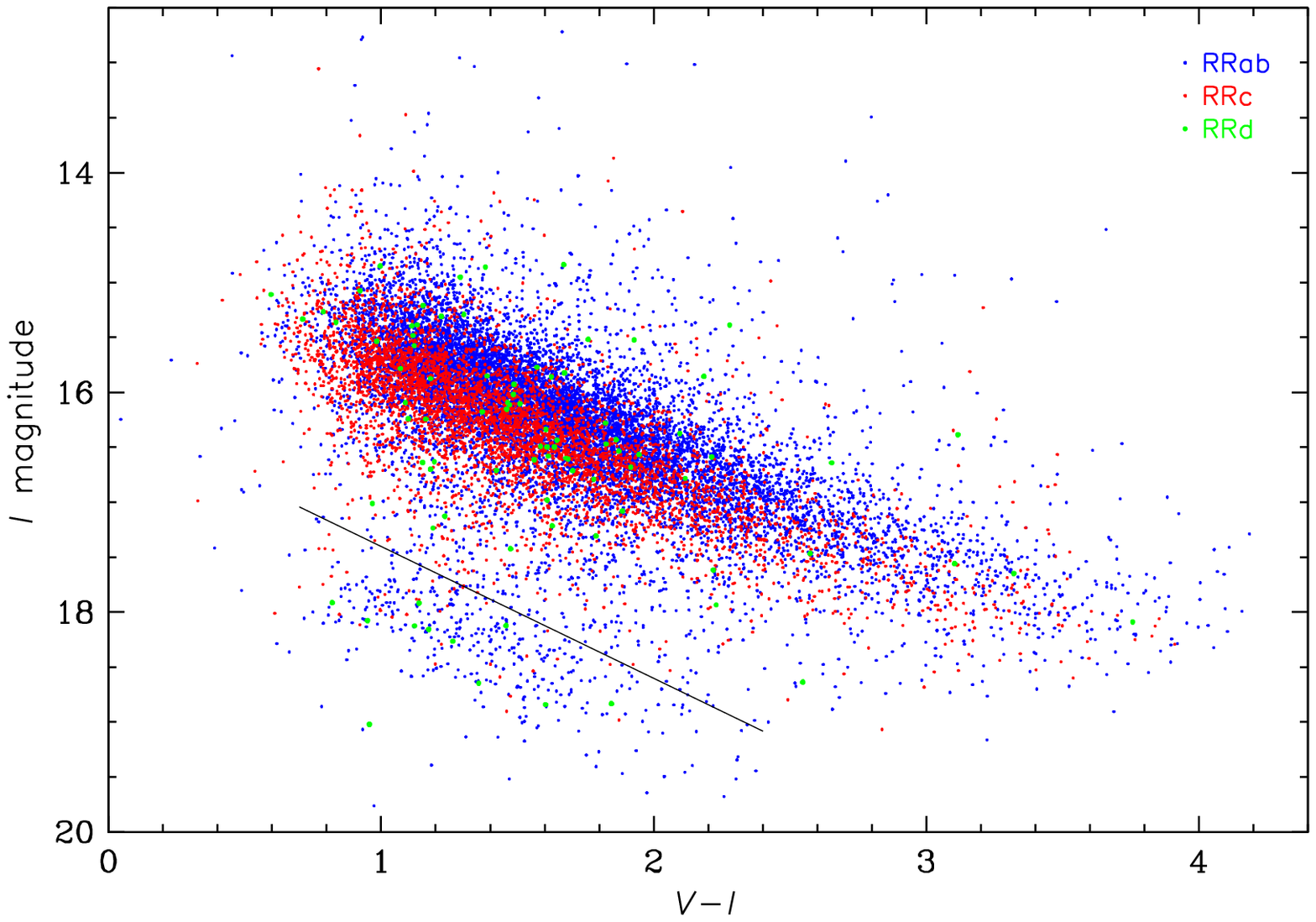}}
\FigCap{Color--magnitude diagram for RR~Lyr stars toward the Galactic
bulge. Different colors correspond to types of stars as shown in Fig.~1.
RR~Lyr stars below the black line have been recognized as members of the
Sgr~dSph.}
\end{figure}

The selection and classification of variable stars based primarily on the
morphology of light curves. Short-period variable stars were divided into
two groups: pulsating stars and much more numerous group of eclipsing and
ellipsoidal binaries, which will be published in a future part of the
OIII-CVS. The vast majority of pulsating stars were categorized as RR~Lyr
stars, only a small fraction was classified as Cepheids and $\delta$~Sct
stars due to their characteristic light curve shapes or period ratios in
double-mode pulsators. Note, that our catalog may still contain a small
fraction of $\delta$~Sct stars, which are difficult to distinguish from
short-period RR~Lyr variables, when their absolute magnitudes are not {\it
a priori} known.

It was relatively easy to discriminate RRab stars from overtone pulsators
and other types of variable stars, since fundamental-mode RR~Lyr stars have
characteristic, asymmetric light curves. The correct classification was
more problematic in the case of the overtone pulsators (RRc stars), as
their light curves are much more sinusoidal and may be confused with W~UMa,
ellipsoidal, rotating, etc. variable stars. In this catalog we classified
as RRc stars only those objects, which reveal detectable asymmetry of their
light curves. This affects the completeness of the RRc list among the
fainter stars. In difficult cases we took into account the position of a
star in the period--Fourier coefficients (Fig.~1), period--amplitude
(Fig.~2), color--magnitude diagrams (Fig.~3), and a ratio of amplitudes in
the {\it V-} and {\it I}-bands (when the number of observing points in the
{\it V} band was high enough to determine the amplitude in this
band). However, the classification of about one hundred objects in our
catalog remains uncertain. Information about these stars can be found in
the remarks of the catalog.

In contrast to the OGLE-III catalogs of RR~Lyr stars in the Magellanic
Clouds (Papers~I and II), we have not distinguished between RRc and RRe
stars, \ie the the first- and potential second-overtone pulsators. Despite
the fact that RR~Lyr stars in the bulge are closer than in the Magellanic
Clouds, so the quality of the photometry is better, we have not noticed any
natural boundary between RRc and RRe variables.

\Subsection{Multi-Periodic Variables}
RR~Lyr stars pulsating simultaneously in two radial modes (RRd stars) are
very rare in the Galactic bulge (Moskalik and Poretti 2003, Mizerski
2003). Only five objects of this type have been known to date in this
region of the Milky Way. Our search for multiperiodic RR~Lyr stars has been
carried out in two ways. First, we used the database of periods measured
for all stars observed by OGLE in the bulge. We selected and visually
inspected light curves with periods and period ratios characteristic for
the previously known RRd stars, \ie with longer periods in the range
0.42--0.6~days and the shorter-to-longer period ratios between 0.74 and
0.75. Second, we performed a search for secondary periods in the previously
selected set of RR~Lyr stars. Each light curve was fitted with the Fourier
sum with the number of elements that minimizes the $\chi^2$ per degree of
freedom. Then, the function was subtracted from the light curve and the
search for additional periodicities was performed on the residual data.
\begin{figure}[htb]
\hglue-4mm{\includegraphics[width=13.3cm, bb=10 225 585 755]{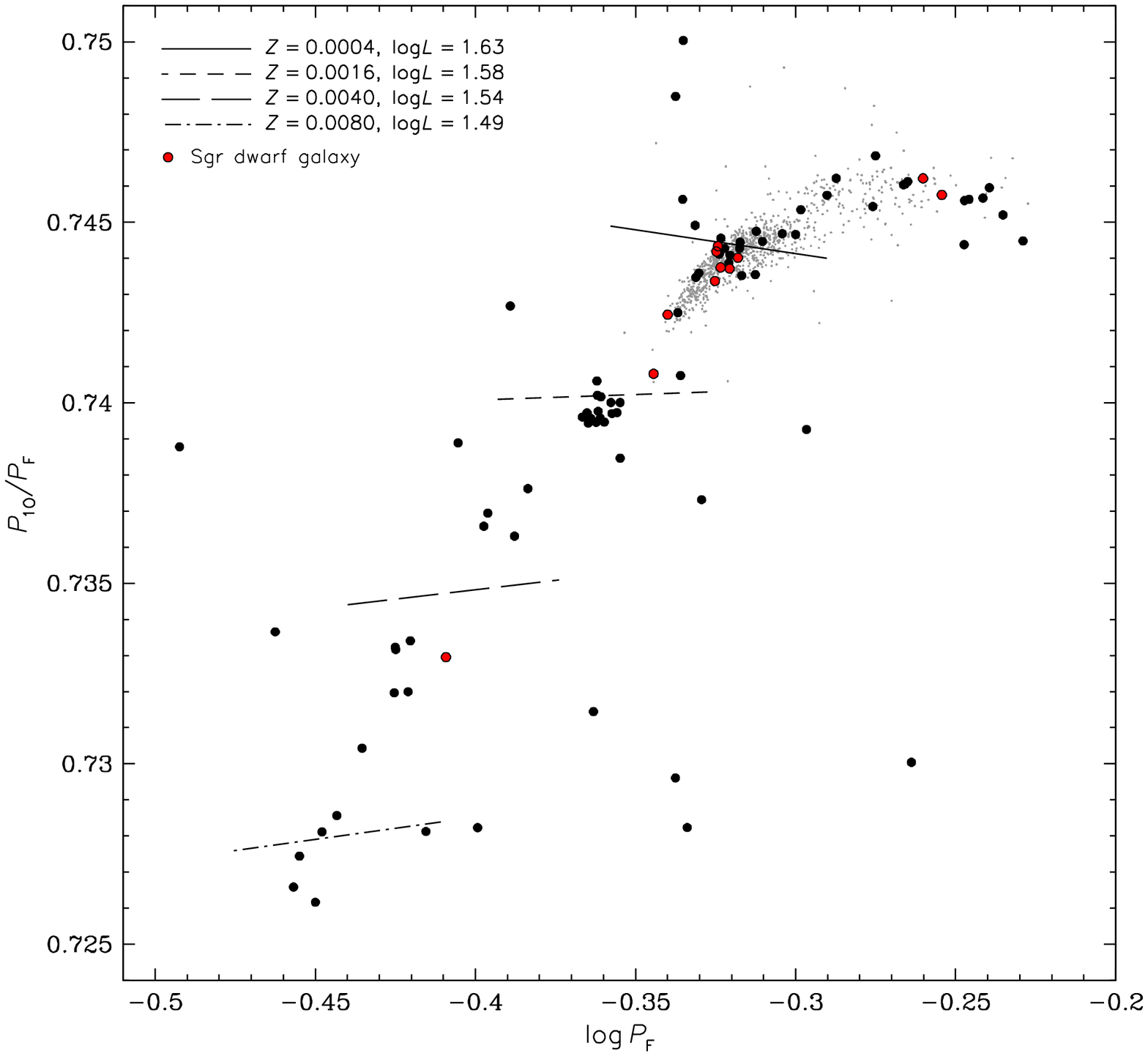}}
\FigCap{Petersen diagram for RRd stars toward the Galactic bulge. Red
symbols represent RRd stars in the Sgr~dSph. Small grey dots show RRd stars
from the LMC (Paper~I). The short segments depict calculated values for
selected models covering central part of the instability strip for a mass
of $M\approx0.6~\MS$. The metal abundance parameter, $Z$, and luminosity,
$L$, are given in the legend.}
\end{figure}

The latter method revealed, somewhat surprisingly, that the well known
sequence in the Petersen diagram (\ie the diagram of the period ratios \vs
the longer periods) has its continuation toward shorter periods and smaller
period ratios. Fig.~4 shows the Petersen diagram for RRd stars in the
bulge. For comparison we plotted 986 RRd stars detected in the LMC
(Paper~I). Though the total number of the RRd in the Galactic bulge is by
two orders of magnitude lower than in the LMC, yet the range of the period
ratios is considerably wider. This appears strange but in part may be
explained by the difference in metal abundance between these two
environments. Selected results of our calculations shown as the segments in
Fig.~4 demonstrate that models of high metal abundance account for the low
values of the period ratios in the bulge RRd stars. In Section 6.2 we
discuss application of RRd stars as a probe of metallicity.

In total, we identified 91 RRd stars (0.5\% of the whole sample of RR~Lyr
stars), confirming very low incident rate of these stars in the bulge. In
the LMC (Paper~I) RRd stars constitute almost 4\% of the total sample of
RR~Lyr stars, while in the SMC (Paper~II) more than 10\% of RR~Lyr stars
are double-mode pulsators. Ten of the 91 detected RRd stars in the bulge
are brighter than typical RR~Lyr stars in the Galactic center, so they are
likely located in front of the bulge. Further 20 RRd stars are distinctly
fainter than bulge RR~Lyr variables, so they are located behind the
bulge. Among them, 11 RRd stars most likely belong to the Sgr~dSph.

In the Petersen diagram one should notice a compact group of 16 RRd stars
around $P_{\rm 1O}/P_{\rm F}{\approx}0.74$ and $\log{P_{\rm
F}}{\approx}{-}0.36$. All these objects have the overtone mode much stronger
than the fundamental one, with the amplitude ratio $A_{\rm 1O}/A_{\rm
F}{>}2.5$. We believe that the similarity of these stars is not by
accident, and probably these objects are relicts of a disrupted dwarf
galaxy or stellar cluster.

During the search for double-mode RR~Lyr stars we found a significant
number of objects with the secondary periods very close to the primary
periods. Such a behavior may be related to the Blazhko effect (Bla{\v z}ko
1907) or changes of the primary period. Long-term OGLE photometry offers an
opportunity to study both phenomena. Using the methods described by Poleski
(2008), we initially selected RR~Lyr stars with detectable rates of period
change. We performed this search only among objects with high quality
photometry (brighter than 16.5~mag in {\it I}) covering a time baseline
longer than 2000~days. As a result we obtained incident rates of RR~Lyr
stars with variable periods. RRab stars that change their periods are
relatively rare and constitute less than 4\% of the total population.
Variable stars with unstable periods are much more common among overtone
pulsators, and reach 38\% of all RRc stars. In this group changes of
periods are more frequent among longer-period variables. About 75\% of RRc
stars with periods in the range of 0.35--0.45~days show detectable rates of
period changes.

Our preliminary analysis confirm a high incident rate of Blazhko-type RRab
stars in the Galactic bulge (Mizerski 2003). At least 30\% of RRab stars
with high-quality photometry exhibit closely-spaced secondary frequencies.
Among RRc stars Blazhko variables constitute about 8\% of the whole
population (excluding stars with variable periods). Fig.~5 presents example
light curves with exceptionally long Blazhko periods (up to about
3000~days) or with large amplitude variations.
\begin{figure}[htb]
\hglue-1mm{\includegraphics[width=13cm]{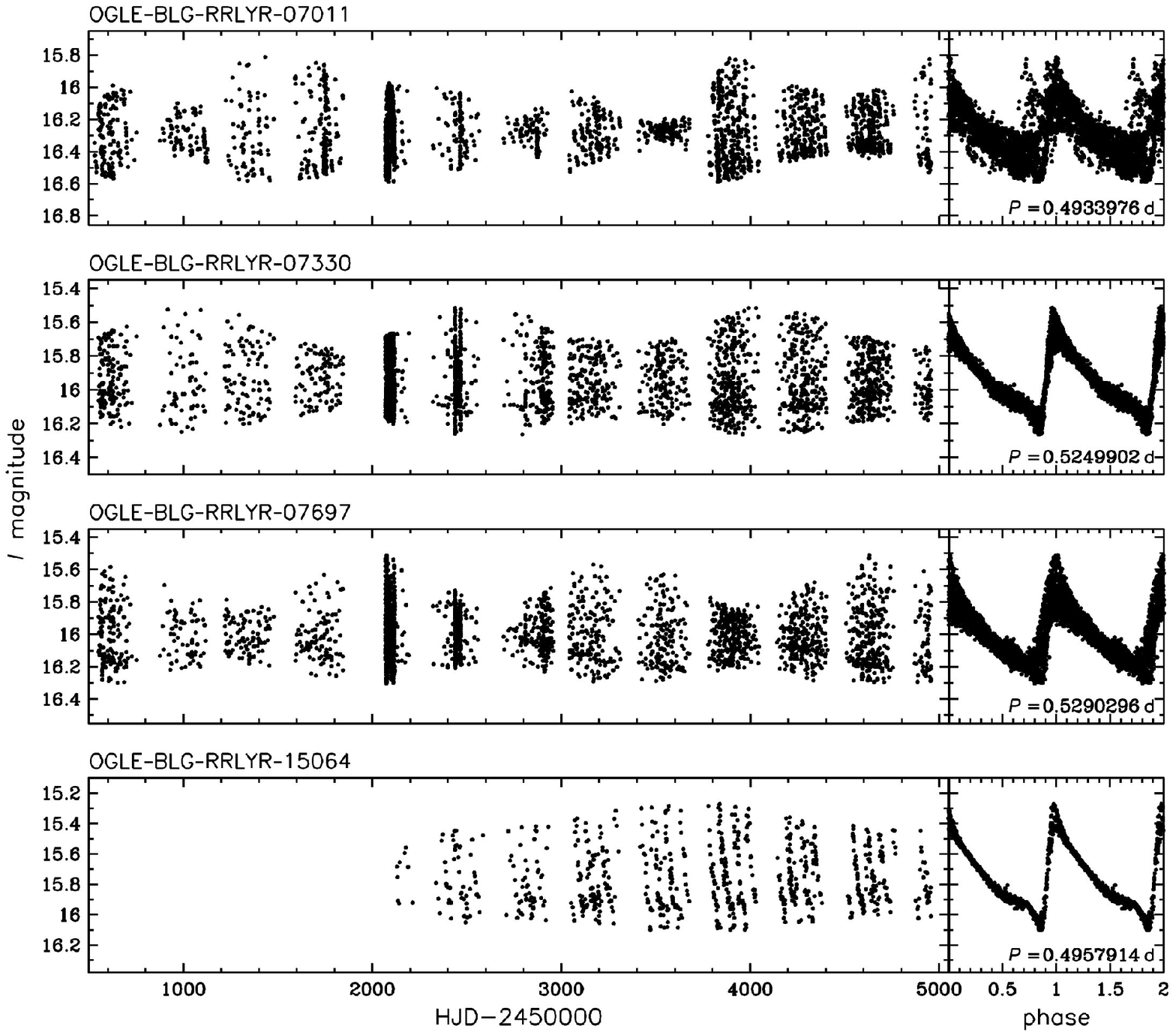}}
\FigCap{Light curves of four RR~Lyr stars with the Blazhko effect. {\it
Left panels}: unfolded OGLE-II (if available) and OGLE-III {\it I}-band
light curves. {\it Right panels}: the same light curves folded with the
pulsation periods.}
\end{figure}

Despite many years of efforts, there is not even one case of RR~Lyr star in
a binary system known today. During the search for the secondary periods we
paid particular attention to the eclipsing variations superimposed on the
pulsation light curves. In the bulge we detected one promising candidate
for an RR~Lyr star in an eclipsing binary system. The light curve of
OGLE-BLG-RRLYR-02792 is plotted in Fig.~6. The original {\it I}-band
photometry folded with the pulsation period is shown in the left panel,
while the right panel shows the eclipsing light curve after subtracting the
RR~Lyr component. Further spectroscopic observations would confirm or
exclude the possibility that we detected an RR~Lyr star being a member of
the binary system. It is interesting to note that we found very similar (in
the sense of the pulsation and orbital periods and the light curve shapes)
case of an RR~Lyr star with eclipsing modulation in the LMC
(Paper~I). Besides, we identified in the Galactic bulge three additional
RR~Lyr stars (OGLE-BLG-RRLYR-03539, \hbox{-09197}, -11361) that exhibited one
eclipsing-like fading during the whole time span covered by the OGLE-III
observations. These objects will be monitored during the OGLE-IV phase.
\begin{figure}[htb]
\hglue-1mm{\includegraphics[width=15cm, bb=100 600 585 755]{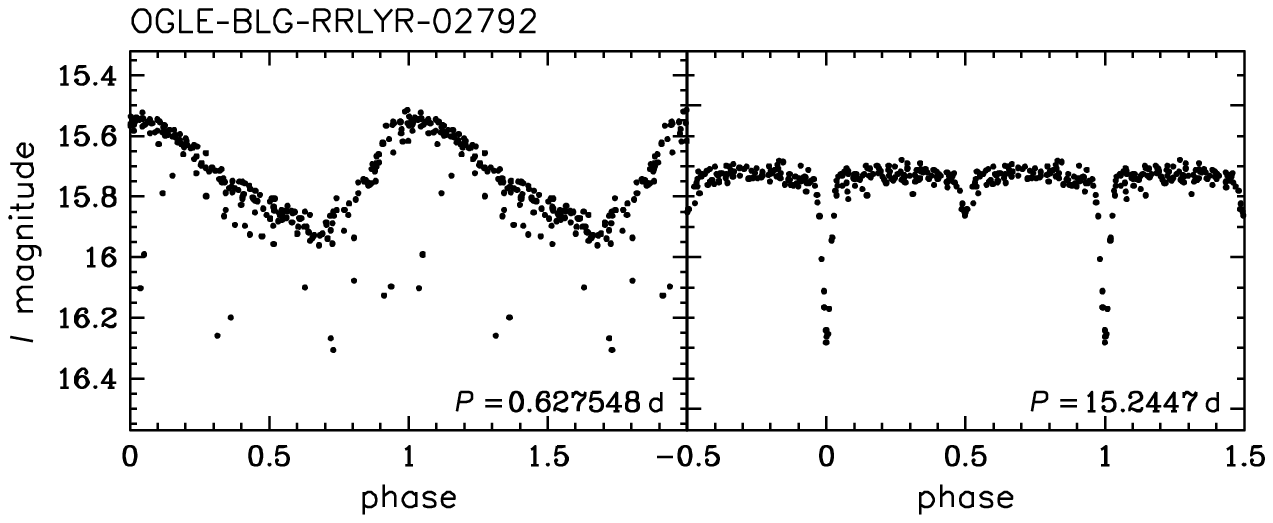}}
\FigCap{Light curve of the RR~Lyr star with additional eclipsing
variability. {\it Left panel}: the original photometric data folded with
the pulsation period. {\it Right panel}: eclipsing light curve after
subtracting the RR~Lyr component.}
\end{figure}

\Section{Catalog of RR~Lyr Stars Toward the Galactic Bulge}
The OGLE-III Catalog of RR~Lyr Stars in the Galactic Bulge consists of
16\,836 objects, of which 11\,756 have been classified as RRab, 4989 as RRc
and 91 as RRd stars. 394 objects in our catalog (343 RRab, 40 RRc and 11
RRd stars) likely belong to the Sgr~dSph. The list of all stars, their
identifications with the previously published catalogs, basic parameters,
time-series {\it I-} and {\it V}-band photometry and finding charts are
available only in electronic form {\it via} FTP site or WWW interface:
\begin{center}
{\it http://ogle.astrouw.edu.pl/} \\ {\it
ftp://ftp.astrouw.edu.pl/ogle/ogle3/OIII-CVS/blg/rrlyr/}\\
\end{center}

\begin{figure}[p]
\hglue-4mm{\includegraphics[width=13.5cm]{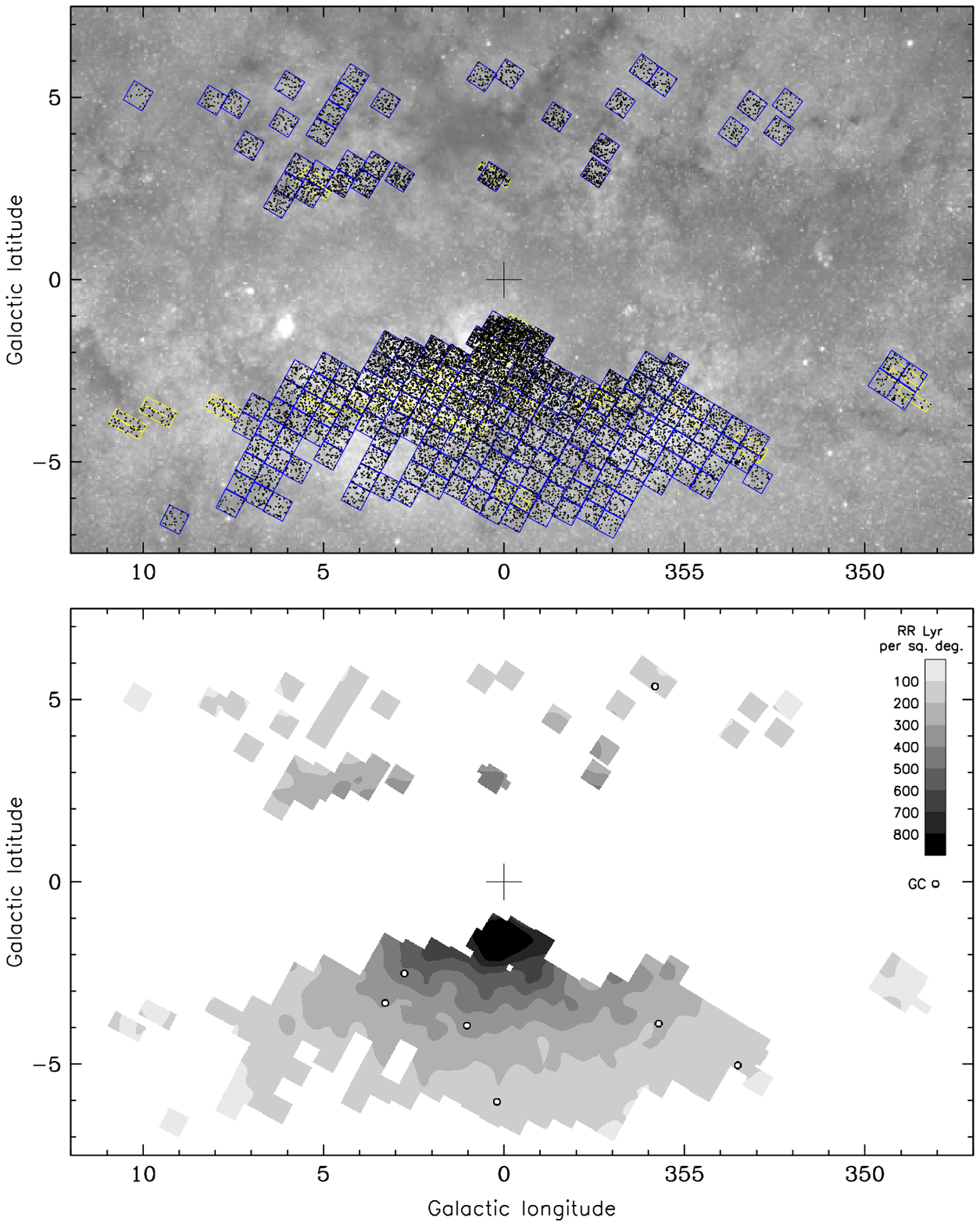}}
\vskip3mm
\FigCap{{\it Upper panel}: spatial distribution of RR~Lyr stars toward the
Galactic bulge. The background image of the bulge originates from the Axel
Mellinger's Milky Way Panorama (Mellinger 2009). Yellow and blue contours
show OGLE-II and OGLE-III fields with the number of observations exceeding
30. {\it Lower panel}: surface density map of RR~Lyr stars toward the
Galactic bulge obtained by blurring the upper map with the Gaussian
function. White circles show positions of globular clusters which contain
RR~Lyr stars.}
\end{figure}
The file {\sf ident.dat} at the FTP site lists all RR~Lyr stars with their
coordinates and identifications in various databases. Designations of
objects in this catalog follow the scheme presented in the previous parts
of the OIII-CVS -- stars are named with the symbols OGLE-BLG-RRLYR-NNNNN,
where NNNNN is a five-digit consecutive number. Objects are arranged
according to increasing right ascension. The subsequent columns in the file
{\sf ident.dat} give: star designation, OGLE-III field and internal
database number (consistent with the photometric maps of the bulge by
Szymañski \etal in preparation), mode of pulsation (RRab, RRc, RRd),
J2000.0 right ascension and declination, cross-identifications with the
OGLE-II photometric database (Szymañski 2005), cross-identifications with
the MACHO catalog of RR~Lyr stars in the bulge (Kunder \etal 2008) and
cross-identifications with the General Catalogue of Variable Stars (GCVS,
Kholopov \etal 1985).

Observational parameters of the RR~Lyr stars -- intensity-averaged {\it I}
and {\it V} magnitudes, periods with uncertainties (derived with the {\sc
Tatry} code of Schwarzenberg-Czerny 1996), peak-to-peak {\it I}-band
amplitudes, epoch of maximum light and Fourier parameters $R_{21}$,
$\phi_{21}$, $R_{31}$, $\phi_{31}$ (Simon and Lee 1981) derived for {\it
I}-band light curves -- are provided in the files {\sf RRab.dat}, {\sf
RRc.dat}, and {\sf RRd.dat}. The latter file gives relevant information
about both periodicities of the double-mode stars. When the number of
observing points in the {\it V}-band was less than 20, we derived mean {\it
V} magnitude by fitting a template light curve, which was obtained from
scaled and shifted {\it I}-band light curve. Additional information on some
objects (\eg additional periods, uncertain classification, proper motion,
etc.) can be found in the file {\sf remarks.txt}. The OGLE-II and OGLE-III
multi-epoch {\it VI} photometry can be downloaded from the directory
{\sf phot/}. Finding charts for each star are stored in the directory
{\sf fcharts/}. These are $60\arcs\times60\arcs$ subframes of the
{\it I}-band DIA reference images, oriented with N up and E to the left.

A spatial distribution of RR~Lyr stars from our catalog is presented in
Fig.~7. The upper panel shows individual stars plotted on the background
image originated from the Axel Mellinger's Milky Way Panorama (Mellinger
2009). Contours of the OGLE-II and OGLE-III fields (only those with the
number of observing points larger than 30) are also plotted in Fig.~7. The
bottom panel in Fig.~7 presents a surface density map obtained by the
convolution of the upper distribution with the Gaussian function. The
strong concentration of RR~Lyr stars toward the Galaxy center is well
visible.

\Section{Completeness of the Catalog}
The RR~Lyr stars in our catalog cover practically the entire range of
magnitudes ($13<I<20.5$~mag) that may be detected with the OGLE data. We
expect that the completeness of the catalog strongly depends on the
brightness of stars, amplitudes, shape of the light curves, stars
surrounding individual objects and number of observing points. In order to
test and improve the completeness of our catalog we compared our sample
with the MACHO and OGLE-II catalogs of RR~Lyr stars in the Milky Way center
and with the previous identifications collected by the GCVS.

The largest list of RR~Lyr stars in the Galactic bulge hitherto published
is the catalog of RRab stars by Kunder \etal (2008) compiled from the MACHO
photometry. Among 2114 MACHO RR~Lyr stars that are covered by the OGLE
fields, we found counterparts for 2087 (98.7\%) objects in the preliminary
version of our catalog. This result may be regarded as the upper limit for
our catalog completeness, and it is valid only for brighter
fundamental-mode RR~Lyr stars. We carefully checked the missing 27 objects
and noticed that 13 of them were located close to bright, saturated stars
and were masked during the reduction process. We included these RR~Lyr
stars in the final version of the catalog providing their {\sc DoPhot}
photometry. Most of the remaining missing RR~Lyr stars were affected by a
small number ($<30$) of observing points, usually due to their location at
the edge of the OGLE fields. When it was possible, we supplemented our
catalog with these objects.

\begin{figure}[p]
\hglue-7mm{\includegraphics[width=15.3cm, bb=10 50 585 705]{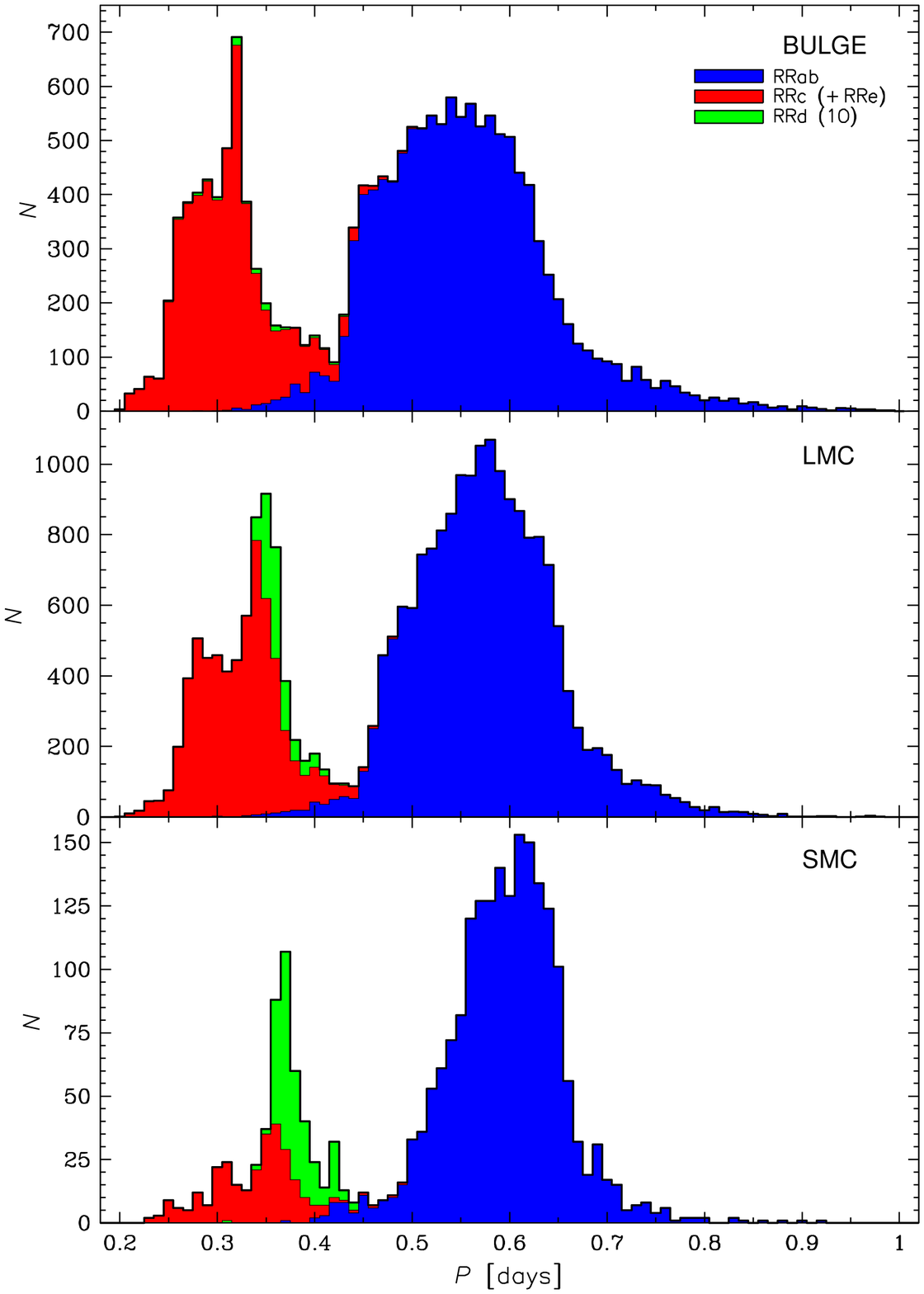}}
\FigCap{Period distribution of RR~Lyr stars in the Galactic bulge, LMC and
SMC. Each color represents different type of pulsators. Blue regions show
RRab stars, red -- RRc (+RRe) stars and green -- the first-overtone period
of RRd stars. The width of bins is 0.01~day.}
\end{figure}
We also cross-identified our catalog with the list of 1888 RRab stars
detected by Collinge \etal (2006) in the OGLE-II fields. We missed only one
object -- a blended star, and thus with reduced amplitude. An independent
test of the completeness of our catalog was the search for RR~Lyr stars
carried out by us in the OGLE-II fields using generally the same methods as
in the OGLE-III fields. In this way we extended our catalog by more than
400 RR~Lyr stars, mostly in the regions monitored by the OGLE-II survey,
and not covered by the OGLE-III fields, or with number of points collected
during the OGLE-III phase smaller than 30. Five of these newly detected
RR~Lyr stars could potentially be identified on the basis of the OGLE-III
data only, but were overlooked at the first stage of the search. Most of
them were faint RRc variables with nearly sinusoidal light curves and
initially were categorized as close binaries.

Among stars classified as RR~Lyr variables in the GCVS (Kholopov \etal
1985), 403 objects can potentially be found in the OGLE bulge fields. We
successfully identified 371 of them in our catalog. Half of the missing
stars have no information about periods in the GCVS, and we could not
properly identify them in the dense bulge fields. From the remaining stars,
seven turned out to be eclipsing binaries, for further seven stars we were
not able to identify objects with similar periods close to their
coordinates and only two objects were identified as saturated RR~Lyr stars.

In summary, our catalog of RR~Lyr stars in the Galactic bulge is close to
be complete, especially for RRab stars. For RRc stars the completeness
depends on the light curve shape (sinusoidal variable stars may be missed)
and brightness (among the faintest RR~Lyr stars in our catalog the incident
rate of RRc stars is artificially reduced due to the problems with
classification). One should remember that the optical OGLE photometry is
not able to penetrate regions highly obscured by the interstellar
medium. The faintest stars observed by OGLE have {\it I}-band magnitudes of
about 20.5~mag. For this reason the spatial density of the RR~Lyr is
underestimated in the narrow area close to the Galaxy center (see Fig.~7).
These regions will be observed in the near-infrared domain by the VISTA
Variables in the Via Lactea (VVV) survey (Minniti \etal 2010). Our catalog
is also incomplete in the very cores of globular clusters, due to the
extreme spatial density of stars in these regions.

\Section{Discussion}
\Subsection{Period Distribution}
The distribution of periods of RR~Lyr stars is a powerful tool for studying
properties of the oldest stellar population. It is well known that average
periods are correlated with the metallicity of RR~Lyr stars, or more
specifically, longer-period variables are generally more metal-poor. Fig.~8
displays the histograms of periods of RR~Lyr stars from the bulge, LMC
(Paper~I) and SMC (Paper~II). Each bin was proportionally divided among
different modes of pulsation and presented in different colors. RRc and RRe
stars from the Magellanic Clouds were combined and marked with the same
(red) color in Fig.~8.

It is clear that RR~Lyr stars in the Galactic bulge have on average shorter
periods than in the Magellanic Clouds. The mean period of RRab stars in the
bulge is 0.556~days, which is exactly 0.02~days shorter than in the LMC
(0.576~days) and 0.04~days shorter than in the SMC (0.596~days). The
difference between these RRab populations is larger, when comparing the
most frequent (modal) periods: 0.54, 0.58 and 0.62~days for the bulge, LMC
and SMC, respectively. Also the overtone RR~Lyr variables have shorter mean
periods in the more metal-rich environments: 0.310~days (mode: 0.32~days)
in the bulge, 0.323~days (mode: 0.34~days) in the LMC (merging together RRc
and RRe stars), and 0.338~days (mode: 0.37~days) in the SMC.

The existence of the second-overtone pulsators among RR~Lyr stars (RRe) is
a matter of debate. There is no doubt that the overtone RR~Lyr variables in
the Galaxy center show two maxima in the period distribution, although the
short-period peak is not as prominent as in the LMC and SMC. Moreover, the
``RRe peak'' does not follow the rule defined by the ``RRab'' and ``RRc
peaks'', \ie the bulge RRe stars do not have shorter periods than the LMC
ones. The local maximum in the period distribution for the short-period
overtone RR~Lyr stars is at 0.29~days for the bulge, 0.28~days for the LMC
and 0.31~days for the SMC. The origin of this additional peak in the period
distribution of RR~Lyr stars remains a mystery.

\Subsection{The RRd Stars}
\begin{figure}[t]
\hglue-1mm{\includegraphics[width=12.8cm]{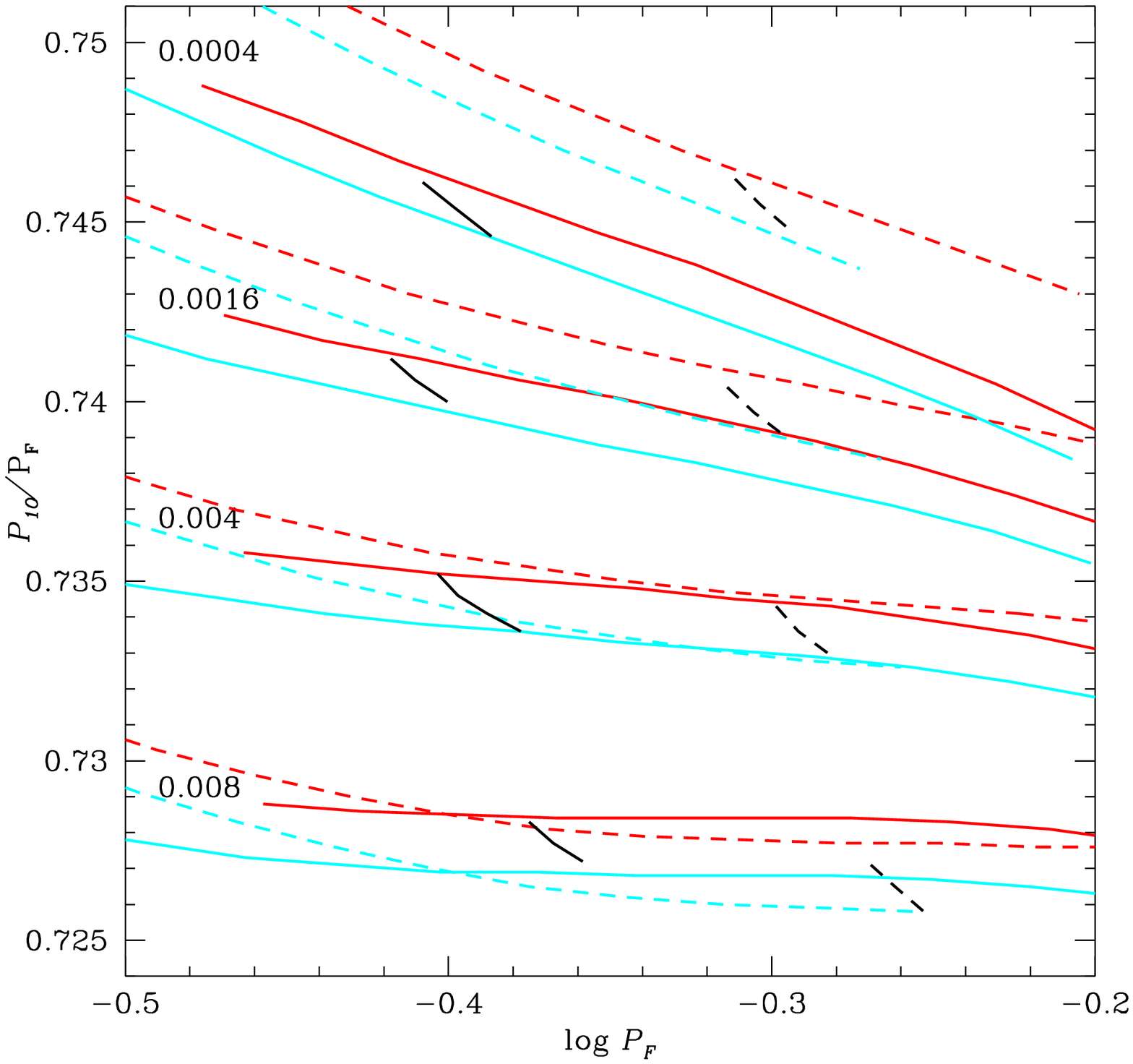}}
\FigCap{RR~Lyr star models in the Petersen diagram. At each four
indicated values of the metallicity parameter, $Z=0.0004$,
0.0016, 0.004, and 0.008 (the respective [Fe/H] values are $-1.67$,
$-1.07$, $-0.67$, and $-0.36$) there are four lines. The solid lines
correspond to $M=0.7~\MS$ and dashed to $M=0.55~\MS$. The cyan and red
colors correspond to the hot and cool boundaries, respectively, of the
adopted effective temperature range. Luminosity varies along these lines
from $\log{(L/\LS)}=1.42$ to $\log{(L/\LS)}=1.75$. The short black segments
show the loci of the frequencies $\nu_{\rm 1O}=0.5(\nu_{\rm F}+\nu_{\rm
2O})$ within the temperature range.}
\end{figure}

The segments shown in the Petersen diagram (Fig.~4), were selected from our
survey of the linear pulsation of stellar envelope models in the relevant
range of parameters. More results from this survey is shown in Fig.~9. We
considered models with masses and luminosities appropriate for horizontal
branch stars. In the adopted effective temperature range, which is about
300~K wide, first two radial modes are unstable. Comparing these two
figures, we note that with the adopted range of $Z$, almost the whole
observed range is covered. Only some stars lying in the upper right corner
may need $Z<0.0004$ and those in the lower left corner $Z>0.008$.

At the specified $P_{\rm F}$, the period ratio mainly depends on $Z$. A
slight decrease with increasing $T_{\rm eff}$ is seen in the difference
between red and cyan lines. The solid and dashed lines represent different
masses.  Calculated numbers depend somewhat on the adopted heavy element
mixture and source of opacity data, which is more significant (Buchler
2008). In models calculated for Fig.~9 we used the OPAL opacity data.

In any case, to explain the existence of the short period RRd stars in the
Galactic bulge we need to postulate that these objects have metal abundance
much closer to the young population of the LMC, than to RRd stars in this
galaxy. This is acceptable in light of what has been known about
metallicity in the bulge. Let us note that to explain the exceptionally
short-period tail in the distribution of bulge RRab stars seen in Fig.~8,
we also need models with high $Z$ values.

Kunder and Chaboyer (2008), who based their assessment of the RR star
metallicity in the Galactic bulge on light curve data, found a broad range
of the [Fe/H] values extending up to $-0.15$~dex. Our result provides an
independent evidence for existence of high metallicity RR~Lyr stars in the
Galactic bulge.

The high metallicity RR~Lyr stars in the Galactic field have been known for
long time, but still their existence presents a challenge for stellar
evolution theory. There are no satisfactory evolutionary models starting
from ZAMS for metal rich horizontal branch stars. In particular, even with
enhanced mass loss BaSTI tracks (Pietrinferni \etal 2006) calculated with
$Z\gtrsim0.004$ enter the instability strip during the helium phase only,
if the initial mass is less than 0.9~\MS. However, it takes time longer
than the Universe age for such objects to reach this phase of evolution.
Still larger mass loss in the red giant phase than adopted in the BaSTI
tracks is needed. These issues has been contemplated by various authors
(see \eg Catelan 2009). The question why it is more likely to happen in
the bulge than in other environments remains to be answered.

We also do not have explanation for the large disparity in the incident
rate of double mode pulsation between the LMC and the Galactic bulge RR~Lyr
stars, due to insufficient understanding of how such a form of pulsation
arises. This problem in the context of Cepheid pulsation was discussed
recently by Smolec and Moskalik (2010). One effect that they identify as a
possible source of such a pulsational behavior is the $\omega_{\rm
1O}=\frac{1}{2}(\omega_{\rm F}+\omega_{\rm 2O})$ resonance. It may also
play a role in our sample of RRd stars. The segments in Fig.~9 mark
positions where the resonance condition is satisfied exactly. For other
acceptable models the condition is nearly satisfied.  However, only
nonlinear modeling may provide an answer whether this is the actual cause
of the double mode pulsation.

\begin{figure}[htb]
\includegraphics[width=12.7cm]{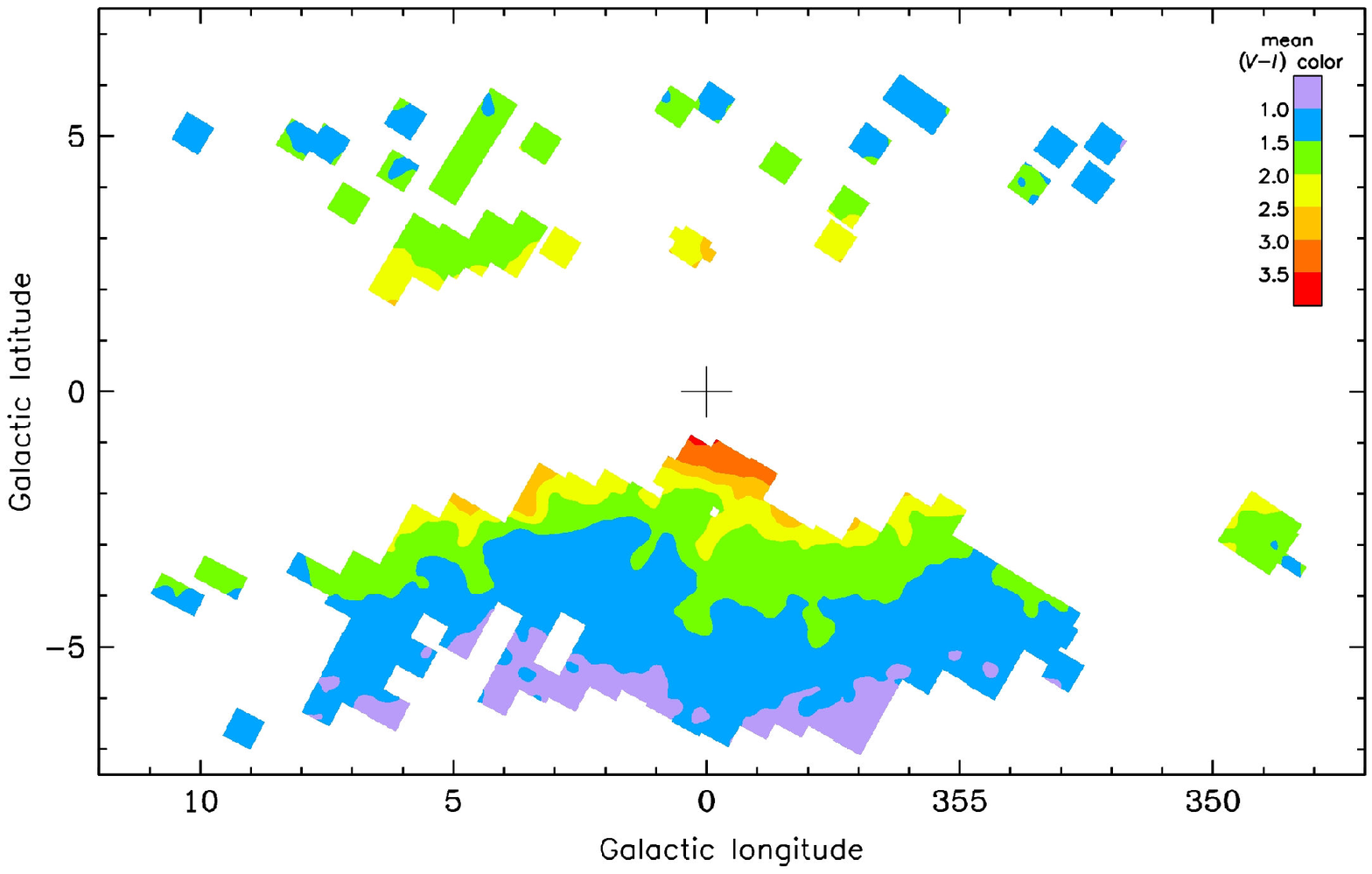}
\FigCap{Spatial distribution of the mean apparent $(V-I)$ colors of RR~Lyr
stars toward the Galactic bulge.}
\end{figure}

\Subsection{Interstellar Extinction}
Since RR~Lyr stars have approximately the same absolute magnitudes and
colors they are excellent indicators of the interstellar extinction, in the
sense of measuring both -- the amount of extinction and the extinction
law. The study of the extinction toward the Milky Way center were
undertaken by Kunder \etal (2008) using RRab stars identified in the MACHO
fields. The map of the interstellar extinction on the basis of our catalog
will be prepared elsewhere. In this paper we present only the map of the
mean apparent $(V-I)$ colors of RR~Lyr stars in the bulge (Fig.~10). Very
large reddening toward the Milky Way center changes the apparent colors of
RR~Lyr stars up to $(V-I)>4$~mag. In the most obscured regions the RR~Lyr
stars are too faint to be observed by the OGLE project in the {\it V}-band,
and only the {\it I}-band light curves are available. In Fig.~10 the lines
of constant mean colors are roughly parallel to equatorial plane of the
Galaxy, which is expected when the absorbing medium is located in the thin
disk in front of the bulge. The deviation from this symmetry visible in
Fig.~10 (Baade's Window) may be related to the inclined barred structure of
the Galaxy center.

\vspace*{6pt}
\Subsection{RR~Lyr Stars in the Sagittarius Dwarf Spheroidal Galaxy}
\vspace*{5pt}
Sagittarius Dwarf Spheroidal Galaxy is a substantially tidally disrupted
satellite of the Milky Way. The galaxy is distributed across much of the
celestial sphere. First RR~Lyr stars in Sgr~dSph were discovered by Mateo
\etal (1995) as a part of the first phase of the OGLE project. During the
next years, the population of known RR~Lyr stars in Sgr~dSph grew
significantly thanks to the studies by Alard (1996), Alcock \etal (1997),
Cseresnjes (2001), Kunder and Chaboyer (2009).

The OGLE-III fields are located at angular distances from 7.6 to 23 degrees
from the globular cluster M~54, which is believed to be the center of
Sgr~dSph. So, our catalog is suitable to study only the outer parts of this
galaxy. The color--magnitude diagram (Fig.~3) clearly shows the sequence of
faint RR~Lyr stars that belong to Sgr~dSph. We separated the Sgr~dSph members from
other RR~Lyr stars by adopting a somewhat arbitrary condition:
$I>1.2(V-I)+16.2$~mag (the line in Fig.~3).

\begin{figure}[htb]
\hglue-8mm{\includegraphics[width=14cm, bb=10 400 585 755]{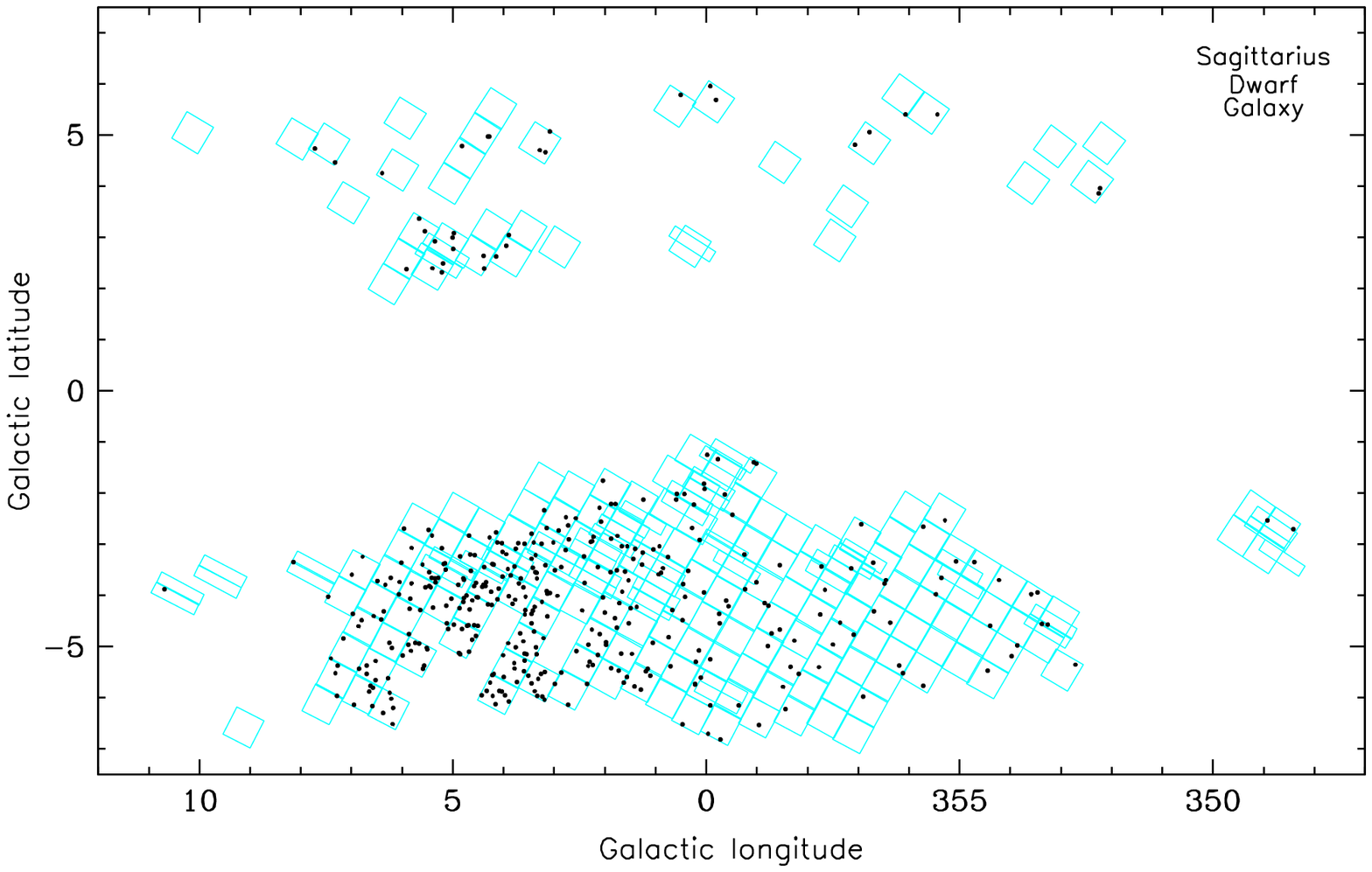}}
\FigCap{Spatial distribution of RR~Lyr stars from the Sagittarius Dwarf
Galaxy.}
\end{figure}
Fig.~11 presents the spatial map of 394 RR~Lyr stars selected in this
way. Our sample is incomplete in the regions close to the Galactic plane,
where the interstellar extinction is very large. It is not surprising since
the color--magnitude diagram (Fig.~3) clearly shows that Sgr~dSph RR~Lyr stars
with apparent colors $V-I>2.4$~mag are too faint to be detected with the
OGLE photometry. In the less obscured area observed by the OGLE project,
the spatial gradient of RR~Lyr stars in the Sgr~dSph is visible.

Since RR~Lyr stars in the Sgr~dSph are close to the detection limit of the
OGLE survey, our sample is incomplete, especially for RRc variables. Most
of the overtone pulsators with symmetric light curves were probably
classified as close binaries due to noisy photometry of such faint stars.

\begin{figure}[p]
\vglue5mm
\hglue-13mm{\includegraphics[width=16cm, bb=10 50 585 715]{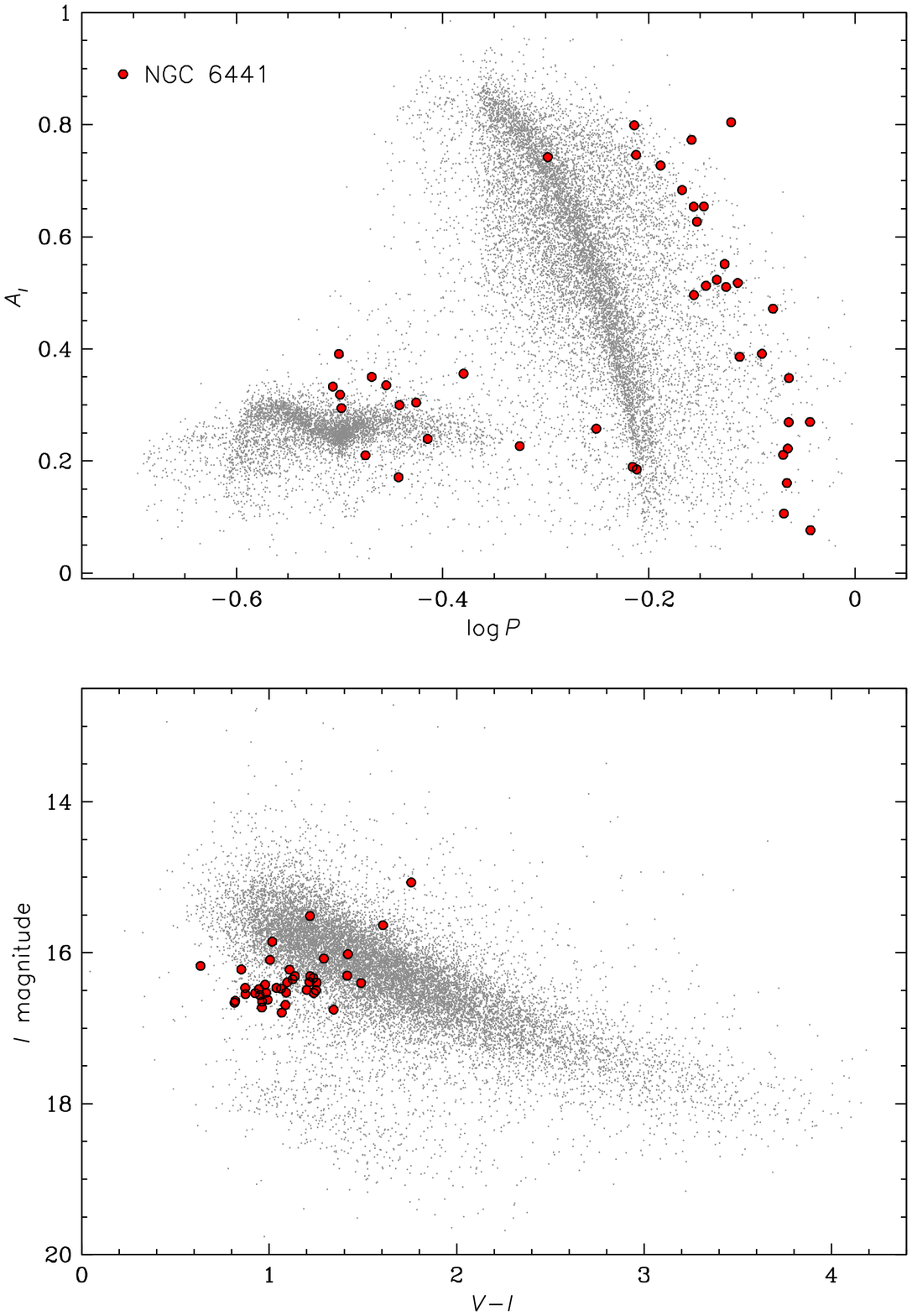}}
\FigCap{Period--amplitude ({\it upper panel}) and color--magnitude ({\it
lower panel}) diagrams for RR~Lyr stars in the globular cluster NGC~6441
(red points). Background grey dots represent other RR~Lyr stars toward the
Galactic bulge.}
\end{figure}

\Subsection{RR~Lyr Stars in Globular Clusters}
The OGLE fields in the Galactic bulge cover ten globular
clusters\footnote{According to the list of Milky Way globular clusters
available at the web page\newline
http://www.seds.org/messier/xtra/supp/mw\_gc.html}. We selected RR~Lyr
stars which lay inside the area outlined by the angular radii of these
clusters. To estimate the number of field RR~Lyr stars, which may be
present by chance within the cluster radii, we counted RR~Lyr stars in the
rings around the clusters (from 1.5 to 2.5 of the cluster radii, but we
checked also other values) and rescaled the number of detected stars to the
area covered in the sky by a cluster. We emphasize that our survey is not
able to detect variable stars in the very cores of the globular clusters.

\MakeTable{
l@{\hspace{8pt}} c@{\hspace{6pt}} c@{\hspace{6pt}} c@{\hspace{8pt}}
c@{\hspace{6pt}} c@{\hspace{6pt}}}{12.5cm}{Globular clusters in the OGLE
fields containing RR~Lyr stars}{\hline \noalign{\vskip3pt}
\multicolumn{1}{c}{Cluster} & RA & Dec & Cluster & $N_{\rm RR}$ & $N_{\rm field RR}$ \\
 \multicolumn{1}{c}{name} & (J2000) & (J2000) & radius [\arcm] & &
 (estimated) \\
\noalign{\vskip3pt}
\hline
\noalign{\vskip3pt}
 NGC~6304 & 17\uph14\upm32\ups & $-29\arcd27\arcm44\arcs$ & 4.0 &  5 &  1.5 \\
 NGC~6441 & 17\uph50\upm13\ups & $-37\arcd03\arcm04\arcs$ & 4.8 & 43 &  2.8 \\
 NGC~6453 & 17\uph50\upm52\ups & $-34\arcd35\arcm55\arcs$ & 3.8 &  6 &  2.0 \\
 Djorg~2  & 18\uph01\upm49\ups & $-27\arcd49\arcm33\arcs$ & 5.0 & 17 & 10.0 \\
 NGC~6522 & 18\uph03\upm34\ups & $-30\arcd02\arcm02\arcs$ & 4.7 & 15 &  6.5 \\
 NGC~6540 & 18\uph06\upm09\ups & $-27\arcd45\arcm55\arcs$ & 0.8 &  3 &  0.5 \\
 NGC~6558 & 18\uph10\upm18\ups & $-31\arcd45\arcm49\arcs$ & 4.2 &  7 &  0.4 \\
\noalign{\vskip3pt}
\hline}

Seven globular cluster which may host RR~Lyr stars are listed in
Table~1. No RR~Lyr stars were found in the following globular clusters:
NGC~6355, NGC~6528 and NGC~6553. Each of the further 3 clusters:
NGC~6304, NGC~6453, NGC~6540, may host up to four RR~Lyr stars, but it
cannot be ruled out that all of the detected pulsators are field
variables. Other four globular clusters observed by OGLE in the bulge --
NGC~6441, Djorg~2, NGC6522 and NGC~6558 -- contain RR~Lyr stars, although
usually only several objects.

An exceptionally rich cluster is NGC~6441, which hosts around 40 RR~Lyr
stars outside its core. RR~Lyr variables in this cluster also have
exceptionally long periods, actually the longest mean periods from all
known globular clusters. Moreover, NGC~6441 together with another globular
cluster, NGC~6388, violates the rule that more metal-rich clusters host
shorter-period RR~Lyr stars. Pritzl \etal (2000) suggested that NGC~6441
and NGC~6388 represent a new, third Oosterhoff group of globular clusters.

Fig.~12 shows the period--amplitude and color--magnitude diagrams for
RR~Lyr stars in NGC~6441 overplotted on other RR~Lyr stars from our
catalog. The $\log{P}$ of the NGC~6441 members is shifted toward longer
periods by about 0.15 compared to the field bulge RRab variables. RR~Lyr
stars in NGC~6441 are significantly fainter than field variables
surrounding the cluster in the sky, confirming the background location of
the cluster with respect to the bulge. A more detailed description of
RR~Lyr stars in globular clusters will be presented in a separate paper.

\Section{Conclusions}
We presented the largest catalog of RR~Lyr stars toward the Galactic bulge
published so far. Our sample is about five times larger than the largest
set of RR~Lyr stars identified in the bulge before. A huge number of
objects distributed over a relatively large area in the central parts of
the Galaxy, high completeness (especially for RRab stars), and excellent
multi-epoch standard photometry gives an opportunity to map a 3D structure
of the bulge, to test the presence of barred distribution among the oldest
population of stars, to explore the earliest history of the Galaxy
formation, and to determine an accurate distance to the Milky Way center.

\Acknow{We are grateful to Z.~Ko³aczkowski, A.~Schwarzenberg-Czerny and
J.~Skowron for providing software which enabled us to prepare this study.

The research leading to these results has received funding from the
European Research Council under the European Community's Seventh Framework
Program\-me (FP7/2007-2013)/ERC grant agreement no. 246678. This work has
been supported by the MNiSW grant no. IP2010 031570 (the Iuventus Plus
program) to P.~Pietrukowicz. The massive period search was performed at the
Interdisciplinary Centre for Mathematical and Computational Modeling of
Warsaw University (ICM), project no.~G32-3. We wish to thank M. Cytowski
for his skilled support.}


\begin{references}
\refitem{Alard, C.}{1996}{\ApJ}{458}{L17}
\refitem{Alard, C., and Lupton, R.H.}{1998}{\ApJ}{503}{325}
\refitem{Alcock, C., \etal (MACHO team)}{1997}{\ApJ}{474}{217}
\refitem{Alcock, C., \etal (MACHO team)}{1998}{\ApJ}{492}{190}
\refitem{Baade, W.}{ 1946}{\PASP}{58}{249}
\refitem{Baade, W.}{ 1951}{Publ. Obs. Univ. Michigan}{10}{7}
\refitem{Blanco, B.M.}{1984}{\AJ}{89}{1836}
\refitem{Bla{\v z}ko, S.}{1907}{\AN}{175}{325}
\refitem{Buchler, J.R.}{2008}{\ApJ}{680}{1412}
\refitem{Catelan, M.}{2009}{IAU Symp.}{258}{209}
\refitem{Collinge, M.J., Sumi, T., and Fabrycky, D.}{2006}{\ApJ}{651}{197}
\refitem{Cseresnjes, P.}{2001}{\AA}{375}{909}
\refitem{Fokker, A.D.}{1951}{Annalen van de Sterrewacht te Leiden}{20}{261}
\refitem{Gaposchkin, S.I.}{1956}{Peremennye Zvezdy}{11}{268}
\refitem{Hartwick, F.D.A., Barlow, D.J., and Hesser, J.E.}{1981}{\AJ}{86}{1044}
\refitem{Kholopov, P.N., \etal}{1985}{~}{~}{``General Catalogue of Variable Stars'', 4th Edition, Nauka Publishing House, Moscow}
\refitem{Kunder, A., and Chaboyer, B.}{2008}{\AJ}{136}{2441}
\refitem{Kunder, A., and Chaboyer, B.}{2009}{\AJ}{137}{4478}
\refitem{Kunder, A., Popowski, P., Cook, K.H., and Chaboyer, B.}{2008}{\AJ}{135}{631}
\refitem{Mateo, M., Kubiak, M., Szymañski, M., Ka³u¿ny, J., Krzemiñski, W., and Udalski, A.}{1995}{\AJ}{110}{1141}
\refitem{Mellinger, A.}{2009}{\PASP}{121}{1180}
\refitem{Minniti, D., \etal}{2010}{New Astronomy}{15}{433}
\refitem{Mizerski, T.}{2003}{\Acta}{53}{307}
\refitem{Moskalik, P., and Poretti, E.}{2003}{\AA}{398}{213}
\refitem{Oosterhoff, P.T., and Horikx, J.A.}{1952}{Annalen van de Sterrewacht te Leiden}{20}{293}
\refitem{Oosterhoff, P.T., Horikx, J.A., and Ponsen, J.}{1954}{Annalen van de Sterrewacht te Leiden}{20}{345}
\refitem{Oosterhoff, P.T., Ponsen, J., and Schuurman, M.C.}{1967}{Bull. Astron. Inst. Neth. Suppl. Ser.}{1}{397}
\refitem{Oosterhoff, P.T., and Ponsen, J.}{1968}{Bull. Astron. Inst. Neth. Suppl. Ser.}{3}{79}
\refitem{Pietrinferni, A., Cassisi, S., Salaris, M., and Castelli, F.} {2006} {\ApJ}{642}{797}
\refitem{Plaut, L.}{1948}{Annalen van de Sterrewacht te Leiden}{20}{3}
\refitem{Plaut, L.}{1973}{\AA}{26}{317}
\refitem{Poleski, R.}{2008}{\Acta}{58}{313}
\refitem{Ponsen, J.}{1955}{Annalen van de Sterrewacht te Leiden}{20}{383}
\refitem{Pritzl, B., Smith, H.A., Catelan, M., and Sweigart, A.V.}{2000}{\ApJ}{530}{L41}
\refitem{Schechter, P.L., Mateo, M., and Saha, A.}{1993}{\PASP}{105}{1342}
\refitem{Schwarzenberg-Czerny, A.}{1996}{\ApJ}{460}{L107}
\refitem{Simon, N.R., and Lee, A.S.}{1981}{\ApJ}{248}{291}
\refitem{Smolec, R. and Moskalik, P.}{2010}{\AA}{524}{40}
\refitem{Soszyñski,~I., Udalski, A., Szymañski,~M.K., Kubiak,~M., Pietrzyñski,~G., Wyrzykowski,~£., Szewczyk,~O., Ulaczyk,~K., and Poleski,~R.}{2009}{\Acta}{59}{1 (Paper~I)}
\refitem{Soszyñski,~I., Udalski, A., Szymañski,~M.K., Kubiak,~M., Pietrzyñski,~G., Wyrzykowski,~£., Ulaczyk,~K., and Poleski,~R.}{2010}{\Acta}{60}{165 (Paper~II)}
\refitem{Swope, H.H.}{1936}{Ann. Harv. Col. Obs.}{90}{207}
\refitem{Swope, H.H.}{1938}{Ann. Harv. Col. Obs.}{90}{231}
\refitem{Szymañski, M.K.}{2005}{\Acta}{55}{43}
\refitem{Udalski, A., Kubiak, M., Szymañski, M., Ka³u¿ny, J., Mateo, M., and Krzemiñski,~W.}{1994}{\Acta}{44}{317}
\refitem{Udalski, A., Szymañski, M., Ka³u¿ny, J., Kubiak, M., Mateo, M., and Krzemiñski,~W.}{1995a}{\Acta}{45}{1}
\refitem{Udalski, A., Olech, A., Szymañski, M., Ka³u¿ny, J., Kubiak, M., Mateo, M., and Krzemiñski,~W.}{1995b}{\Acta}{45}{433}
\refitem{Udalski, A., Olech, A., Szymañski, M., Ka³u¿ny, J., Kubiak, M., Krzemiñski,~W., Mateo, M., and Stanek,~K.Z.}{1996}{\Acta}{46}{51}
\refitem{Udalski, A., Olech, A., Szymañski, M., Ka³u¿ny, J., Kubiak, M., Mateo, M., Krzemiñski,~W., and Stanek,~K.Z.}{1997}{\Acta}{47}{1}
\refitem{Udalski, A.}{2003}{\Acta}{53}{291}
\refitem{Udalski, A., Szymañski, M.K., Soszyñski, I., and Poleski, R.}{2008}{\Acta}{58}{69}
\refitem{van Gent, H.}{1932}{Bull. Astron. Inst. Neth.}{6}{163}
\refitem{van Gent, H.}{1933}{Bull. Astron. Inst. Neth.}{7}{21}
\refitem{Wo¼niak, P.R.}{2000}{\Acta}{50}{421}
\end{references}
\end{document}